\documentclass[ALICE,manyauthors]{cernphprep}
\usepackage[comma,square,numbers,sort&compress]{natbib}
\usepackage[T1]{fontenc}
\usepackage{hyperref}
\usepackage{lineno}
\usepackage{xcolor}
\usepackage{xspace}
\usepackage{multirow}
\usepackage{makecell}
\usepackage{booktabs}
\usepackage{upgreek}
\usepackage{comment}
\usepackage{orcidlink}
\begin{document}
%%%%%%%%%%%%%%%%%%%%%%%%%%%%%%%%%%%%%%%%%%%%%%%%%%
% These are some new commands that may be useful 
% for paper writing in general. If other newcommands
% are needed for your specific paper, please feel 
% free to add here. 
%
% The currently available commands are organized in: 
% 1) Systems
% 2) Operators
% 3) Quantities
% 4) Energies and units
% 5) Detectors
% 6) particle species 
%%%%%%%%%%%%%%%%%%%%%%%%%%%%%%%%%%%%%%%%%%%%%%%%%%

% 1) SYSTEMS 
\newcommand{\ee}               {\mathrm{e^+e^-}} 
\newcommand{\pp}               {\mathrm{pp}}
\newcommand{\ppbar}            {\mathrm{p\overline{p}}}
\newcommand{\pPb}              {\mathrm{p--Pb}}
\newcommand{\PbPb}             {\mathrm{Pb--Pb}}

% 2) OPERATORS
\newcommand{\av}[1]            {\left\langle #1 \right\rangle}

% 3) QUANTITIES 
\newcommand{\s}                {\sqrt{s}}
\newcommand{\sqrtsNN}          {\sqrt{s_\mathrm{NN}}}
\newcommand{\Npart}            {N_\mathrm{part}}
\newcommand{\Ncoll}            {N_\mathrm{coll}}
\newcommand{\RpPb}             {R_\mathrm{pPb}}
\newcommand{\RAA}              {R_\mathrm{AA}}
\newcommand{\TAA}              {T_\mathrm{AA}}
\newcommand{\pt}               {p_\mathrm{T}}
\newcommand{\mt}               {m_\mathrm{T}}
\newcommand{\de}               {\mathrm{d}}
\newcommand{\dEdx}             {\de E/\de x}
\newcommand{\dNdpt}            {\de N/\de\pt}
\newcommand{\dNdptdy}          {\de^{2} N/\de\pt\de y}
\newcommand{\dNdy}             {\de N/\de y}
\newcommand{\mur}              {\mu_\mathrm{R}}
\newcommand{\muf}              {\mu_\mathrm{F}}
\newcommand{\T}                {\mathrm{T}}
\newcommand{\meanpt}           {\langle p_\mathrm{T} \rangle}
\newcommand{\sigmatot}         {\sigma_{\rm tot}}
\newcommand{\fprompt}          {f_\mathrm{prompt}}
\newcommand{\fnonprompt}       {f_\mathrm{non\text{-}prompt}}
\newcommand{\f}[1]             {f_\mathrm{#1}}
\newcommand{\rawY}[1]          {Y_{#1}}
\newcommand{\AccEff}{\ensuremath{(\mathrm{Acc} \times \epsilon)}\xspace}
\newcommand{\effNP}            {(\mathrm{Acc}\times\epsilon)_{\mathrm{non\text{-}prompt}}}
\newcommand{\effP}             {(\mathrm{Acc}\times\epsilon)_{\mathrm{prompt}}}
\newcommand{\Np}               {N_\mathrm{prompt}}
\newcommand{\Nnp}              {N_\mathrm{non\text{-}prompt}}
\newcommand{\qTPC}{q_2^{\rm TPC}}
\newcommand{\qVZEROA}{q_2^{\rm V0A}}
\newcommand{\vtwo}{v_\mathrm{2}}
\newcommand{\vtwotot}{v^\mathrm{tot}_\mathrm{2}}
\newcommand{\vtwosig}{v^\mathrm{sig}_\mathrm{2}}
\newcommand{\vtwobkg}{v^\mathrm{bkg}_\mathrm{2}}
\newcommand{\vtwodplus}{v^\mathrm{D^{+}}_\mathrm{2}}
\newcommand{\vtwoprompt}{v^\mathrm{prompt}_\mathrm{2}}
\newcommand{\vtwofd}{v^\mathrm{non\text{-}prompt}_\mathrm{2}}
\newcommand{\Nsig}{N^\mathrm{sig}}
\newcommand{\Nbkg}{N^\mathrm{bkg}}
\newcommand{\Ndplus}{N^\mathrm{D^{+}}}
\newcommand{\Qtwo}{Q_\mathrm{2}}
\newcommand{\Rtwoflow}{R_{\rm 2}}

% 4) ENERGIES, UNITS
\newcommand{\eV}               {\mathrm{eV}}
\newcommand{\keV}              {\mathrm{keV}}
\newcommand{\MeV}              {\mathrm{MeV}}
\newcommand{\GeV}              {\mathrm{GeV}}
\newcommand{\TeV}              {\mathrm{TeV}}
\newcommand{\ev}               {\mathrm{eV}}
\newcommand{\kev}              {\mathrm{keV}}
\newcommand{\mev}              {\mathrm{MeV}}
\newcommand{\mevc}             {\mathrm{MeV}/c}
\newcommand{\mevcsquared}      {\mathrm{MeV}/c^2}
\newcommand{\gev}              {\mathrm{GeV}}
\newcommand{\gevc}             {\mathrm{GeV}/c}
\newcommand{\tev}              {\mathrm{TeV}}
\newcommand{\fm}               {\mathrm{fm}}
\newcommand{\mm}               {\mathrm{mm}} 
\newcommand{\cm}               {\mathrm{cm}}
\newcommand{\m}                {\mathrm{m}}
\newcommand{\mum}              {\mathrm{\upmu m}}
\newcommand{\ns}               {\mathrm{ns}}
\newcommand{\mrad}             {\mathrm{mrad}}
\newcommand{\mb}               {\mathrm{mb}}
\newcommand{\mub}              {\mathrm{\upmu b}}
\newcommand{\lumi}             {\mathcal{L}_\mathrm{int}}
\newcommand{\nbinv}            {\mathrm{nb^{-1}}}

% 5) DETECTORS 
\newcommand{\ITS}              {\mathrm{ITS}}
\newcommand{\TOF}              {\mathrm{TOF}}
\newcommand{\ZDC}              {\mathrm{ZDC}}
\newcommand{\ZDCs}             {\mathrm{ZDC}}
\newcommand{\ZNA}              {\mathrm{ZNA}}
\newcommand{\ZNC}              {\mathrm{ZNC}}
\newcommand{\SPD}              {\mathrm{SPD}}
\newcommand{\SDD}              {\mathrm{SDD}}
\newcommand{\SSD}              {\mathrm{SSD}}
\newcommand{\TPC}              {\mathrm{TPC}}
\newcommand{\TRD}              {\mathrm{TRD}}
\newcommand{\VZERO}            {\mathrm{V0}}
\newcommand{\VZEROA}           {\mathrm{V0A}}
\newcommand{\VZEROC}           {\mathrm{V0C}}

% 6) PARTICLE SPECIES 
\newcommand{\pip}              {\mathrm{\uppi^{+}}}
\newcommand{\pim}              {\mathrm{\uppi^{-}}}
\newcommand{\kap}              {\mathrm{\rm{K}^{+}}}
\newcommand{\kam}              {\mathrm{\rm{K}^{-}}}
\newcommand{\pbar}             {\mathrm{\rm\overline{p}}}
\newcommand{\kzero}            {\mathrm{K^0_S}}
\newcommand{\lmb}              {\mathrm{\Lambda}}
\newcommand{\almb}             {\mathrm{\overline{\Lambda}}}
\newcommand{\Om}               {\mathrm{\Omega^-}}
\newcommand{\Mo}               {\mathrm{\overline{\Omega}^+}}
\newcommand{\X}                {\mathrm{\Xi^-}}
\newcommand{\Ix}               {\mathrm{\overline{\Xi}^+}}
\newcommand{\Xis}              {\mathrm{\Xi^{\pm}}}
\newcommand{\Oms}              {\mathrm{\Omega^{\pm}}}
\newcommand{\DzerotoKpi}       {\mathrm{D^0 \to K^-\uppi^+}}
\newcommand{\DplustoKpipi}     {\mathrm{D^+\to K^-\uppi^+\uppi^+}}
\newcommand{\DstartoDpi}       {\mathrm{D^{*+} \to \rm D^0 \uppi^+}}
\newcommand{\Dstophipi}        {\mathrm{D_s^+\to \upphi\uppi^+}}
\newcommand{\Dstophipipm}      {\mathrm{D_s^\pm\to \upphi\uppi^\pm}}
\newcommand{\DstophipitoKKpi}  {\mathrm{D_s^+\to \upphi\uppi^+\to K^-K^+\uppi^+}}
\newcommand{\DplustoKKpi}      {\mathrm{D^+\to K^-K^+\uppi^+}}
\newcommand{\phitoKK}          {\mathrm{\upphi\to  K^-K^+}}
\newcommand{\DstoKzerostarK}   {\mathrm{D_s^+\to \overline{K}^{*0} K^+}}
\newcommand{\Dstofzeropi}      {\mathrm{D_s^+\to f_0(980) \uppi^+}}
\newcommand{\fzero}            {\mathrm{f_0(980)}}
\newcommand{\Kzerostar}        {\mathrm{\overline{K}^{*0}}}
\newcommand{\Dzero}            {\mathrm{D^0}}
\newcommand{\Dzerobar}         {\mathrm{\overline{D}\,^0}}
\newcommand{\Dstar}            {\mathrm{D^{*+}}}
\newcommand{\Dstarm}           {\mathrm{D^{*-}}}
\newcommand{\DstarZero}        {\mathrm{D^{*0}}}
\newcommand{\DstarS}           {\mathrm{D_s^{*+}}}
\newcommand{\Dplus}            {\mathrm{D^+}}
\newcommand{\Dminus}           {\mathrm{D^-}}
\newcommand{\Ds}               {\mathrm{D_s^+}}
\newcommand{\Dspm}             {\mathrm{D_s^\pm}}
\newcommand{\Dsstar}           {\mathrm{D_s^{*+}}}
\newcommand{\KKpi}             {\mathrm{K^-K^+\uppi^+}}
\newcommand{\cubar}            {\mathrm{c\bar{u}}}
\newcommand{\cdbar}            {\mathrm{c\bar{d}}}
\newcommand{\ccbar}            {\mathrm{c\overline{c}}}
\newcommand{\bbbar}            {\mathrm{b\overline{b}}}
\newcommand{\Bzero}            {\mathrm{B^0}}
\newcommand{\Bplus}            {\mathrm{B^+}}
\newcommand{\Bzeroplus}        {\mathrm{B^{0,+}}}
\newcommand{\Bs}               {\mathrm{B_s^0}}
\newcommand{\Lambdab}          {\mathrm{\Lambda_b^0}}
\newcommand{\Jpsi}             {\mathrm{J}/\uppsi}
\newcommand{\Vdecay} 	       {\mathrm{V^{0}}}
\newcommand{\bhad}             {\mathrm{H_b}}
\newcommand{\Ztobbbar}         {\mathrm{Z\to b\overline{b}}}
\newcommand{\fctoD}            {f(\mathrm{c}\to\mathrm{D})}
\newcommand{\fbtoB}            {f(\mathrm{b}\to\mathrm{B})}
\newcommand{\fctoHc}           {f(\mathrm{c}\to\mathrm{H_c})}
\newcommand{\fbtoHb}           {f(\mathrm{b}\to\mathrm{H_b})}

%%%%%%%%%%%%%%%  Title page %%%%%%%%%%%%%%%%%%%%%%%%
\begin{titlepage}
% the dates below correspond to CERN approval
% please don't touch: EB chairs will take care
\PHyear{2022}       % required, will be obtained from CERN
\PHnumber{065}      % required, will be obtained from CERN
\PHdate{24 March}  % required, will be obtained from CERN
%%%%%%%%%%%%%%%%%%%%%%%%%%%%%%%%%%%%%%%%%%%%%%%%%%%%

%%% Put your own title + short title here:
\title{Measurement of beauty-strange meson production in Pb--Pb collisions at $\pmb{\sqrtsNN = 5.02~\TeV}$ via non-prompt $\pmb{\Ds}$ mesons}
\ShortTitle{Non-prompt $\Ds$ mesons in Pb--Pb at $\sqrtsNN = 5.02~\TeV$}   % appears on left page headers

%%% Do not change the next lines
\Collaboration{ALICE Collaboration\thanks{See Appendix~\ref{app:collab} for the list of collaboration members}}
\ShortAuthor{ALICE Collaboration} % appears on right page headers, do not change

\begin{abstract}
The production yields of non-prompt $\Ds$ mesons, namely $\Ds$ mesons from beauty-hadron decays, were measured for the first time as a function of the transverse momentum ($\pt$) at midrapidity ($|y|<0.5$) in central and semi-central Pb--Pb collisions at a centre-of-mass energy per nucleon pair $\sqrtsNN=5.02~\tev$ with the ALICE experiment at the LHC.
The $\Ds$ mesons and their charge conjugates were reconstructed from the hadronic decay channel $\Dstophipi$, with $\phitoKK$, in the $4<\pt<36~\GeV/c$ and $2<\pt<24~\GeV/c$ intervals for the 0--10\% and 30--50\% centrality classes, respectively. 
The measured yields of non-prompt $\Ds$ mesons are compared to those of prompt $\Ds$ and non-prompt $\Dzero$ mesons by calculating the ratios of the production yields in Pb--Pb collisions and the nuclear modification factor $\RAA$. The ratio between the $\RAA$ of non-prompt $\Ds$ and prompt $\Ds$ mesons, and that between the $\RAA$ of non-prompt $\Ds$ and non-prompt $\Dzero$ mesons in central Pb--Pb collisions are found to be on average higher than unity in the $4<\pt<12~\GeV/c$ interval with a statistical significance of about $1.6\,\sigma$ and $1.7\,\sigma$, respectively. The measured $\RAA$ ratios are compared with the predictions of theoretical models of heavy-quark transport in a hydrodynamically expanding QGP that incorporate hadronisation via quark recombination.
\end{abstract}
\end{titlepage}

\setcounter{page}{2} %please do not remove this line

%%%%%%%%%%%%%%%%%%%%%%%%%%%%%%%%
% begin main text
%%%%%%%%%%%%%%%%%%%%%%%%%%%%%%%%
\section{Introduction}
\label{sec:intro}
A transition from ordinary nuclear matter to a colour-deconfined medium called quark--gluon plasma (QGP) is predicted to occur at a very high temperature and energy density by quantum chromodynamics (QCD) calculations on the lattice~\cite{Karsch:2006xs,Borsanyi:2020fev,Bazavov:2018mes}, and is supported by several measurements in ultrarelativistic heavy-ion collisions at the SPS, RHIC, and  LHC~\cite{Heinz:2000bk,BRAHMS:2004adc,PHOBOS:2004zne,STAR:2005gfr,PHENIX:2004vcz,ALICE:2018vuu,ALICE:2016ccg,Braun-Munzinger:2015hba}. In such collisions, charm and beauty quarks are mainly produced in hard scattering processes that occur before the formation of the QGP. Hence, they are effective probes of the entire system evolution. While the system undergoes a hydrodynamic expansion, they interact with the medium constituents via elastic~\cite{Thoma:1990fm,Braaten:1991jj,Braaten:1991we} and inelastic~\cite{Baier:1996sk,Gyulassy:1990ye} scatterings. These interactions imply that charm and beauty quarks exchange energy and momentum with the medium constituents, causing high-momentum quarks to lose part of their energy while traversing the QGP. The in-medium energy loss is commonly studied via the measurement of the nuclear modification factor,
\begin{equation}
    \RAA(\pt) = \frac{1}{\langle\TAA\rangle}\times\frac{\de N_\mathrm{AA}/\de \pt}{\de \sigma_\mathrm{pp}/\de\pt},
    \label{eq:RAA}
\end{equation}
where $\de N_\mathrm{AA}/\de\pt$ is the transverse-momentum ($\pt$) differential production yield in nucleus--nucleus collisions, $\de \sigma_\mathrm{pp}/\de\pt$ the $\pt$-differential cross section in proton--proton (pp) collisions, and $\langle\TAA\rangle$ is the average of the nuclear overlap function~\cite{ALICE-PUBLIC-2018-011}. Several measurements of charm and beauty hadrons in Pb--Pb~\cite{ALICE:2021rxa,ALICE:2021kfc,ALICE:2021bib,ALICE:2022tji,CMS:2017qjw,CMS:2017uoy,CMS:2017uuv,CMS:2018bwt,CMS:2018eso,CMS:2019uws,ALICE:2016uid,ATLAS:2018hqe,ATLAS:2021xtw} and Au--Au~\cite{STAR:2018zdy,STAR:2021uzu,PHENIX:2015ynp} collisions show a strong suppression of the production yield at intermediate and high $\pt$ ($\pt>4\text{--}5~\GeV/c$) in heavy-ion collisions compared to pp collisions, suggesting a substantial energy loss of heavy quarks in the QGP. The comparison of the $\RAA$ of light, charm, and beauty hadrons indicates that the energy loss is sensitive to the colour charge and the parton mass. In particular, the $\RAA$ of beauty hadrons is observed to be larger than that of charm hadrons~\cite{ALICE:2022tji,CMS:2017uuv}. For $\pt>5\text{--}6~\GeV/c$, where radiative processes are expected to dominate the energy loss, the smaller suppression is attributed mainly to the so-called “dead cone” effect~\cite{Dokshitzer:1991fd,ALICE:2021aqk}, which suppresses the gluon radiation at angles smaller than $\theta \approx m_\mathrm{Q}/E_\mathrm{Q}$, where $m_\mathrm{Q}$ is the mass of the quark and $E_\mathrm{Q}$ its energy.

Instead, low-$\pt$ heavy quarks experience a “Brownian motion”, which consists of a diffusion process occurring via multiple elastic interactions with low-momentum transfer~\cite{Svetitsky:1987gq}. Owing to the larger mass, beauty quarks diffuse less than charm quarks and have a longer relaxation time, which is expected to be proportional to the quark mass. Measurements of the heavy-flavour hadron production and azimuthal anisotropies can be exploited to constrain the spatial diffusion coefficient $D_s$ via the comparison with theoretical models based on the heavy-quark transport in a hydrodynamically expanding QGP~\cite{ALICE:2020iug,ALICE:2021rxa}. 

A precise description of the hadronisation process in the hot nuclear matter is crucial to understand the transport properties of the QGP~\cite{Rapp:2018qla}. The hadronisation mechanism of low and intermediate-$\pt$ heavy quarks is expected to be sensitive to the presence of a colour-deconfined medium, which could enable hadron formation via quark recombination in addition to the vacuum-like fragmentation. This leads to an enhancement of the production yield of heavy-flavour hadrons with strange-quark content relative to those of non-strange hadrons in Pb--Pb collisions compared to pp collisions, caused by the abundant production of strange--antistrange quark pairs in the QGP~\cite{Rafelski:1982pu,Koch:1986ud,Braun-Munzinger:2015hba}. Recent measurements of the production of prompt $\Ds$ mesons, i.e.~$\Ds$ mesons originating from the charm-quark hadronisation or decays of excited charm-hadron states, by the STAR~\cite{STAR:2021tte} and ALICE~\cite{ALICE:2018lyv,ALICE:2021kfc} Collaborations suggest a relevant role of the recombination mechanism in the charm-quark hadronisation. 
Similar studies in the open-beauty sector, conducted by the CMS Collaboration via the measurement of the $\Bs$-meson production relative to that of $\Bplus$ mesons, show a hint of enhanced production of strange over non-strange mesons~\cite{CMS:2018eso,CMS:2021mzx}. However, no firm conclusions can be drawn within the current uncertainties. Complementary information about the heavy-quark hadronisation in presence of the medium is provided by the measurements of charm baryons and charmonia in heavy-ion collisions~\cite{STAR:2019ank,ALICE:2021bib,CMS:2019uws,ALICE:2016flj,ALICE:2019lga,ALICE:2019nrq}. Recently, the production of heavy-flavour hadrons containing strange quarks was also investigated in high-multiplicity pp collisions~\cite{ALICE:2021npz,LHCb:2022syj}, following the observation of an enhanced production of strange and multi-strange hadrons with increasing charged-particle multiplicity in the light-flavour sector~\cite{ALICE:2016fzo}.

In this Letter, the measurement of the production of $\Ds$ mesons originating from beauty-hadron decays (non-prompt) is reported for central (0--10\%) and semicentral (30--50\%) Pb--Pb collisions at a centre-of-mass energy per nucleon pair $\sqrtsNN=5.02~\TeV$. Non-prompt $\Ds$ mesons provide information about the diffusion and the energy loss of beauty quarks in the QGP. In addition, together with the measurement of non-prompt $\Dzero$ mesons, they have the potential to reveal the beauty-quark hadronisation mechanisms in the QGP, since in pp collisions about 50\% of non-prompt $\Ds$ mesons are produced in $\Bs$ decays~\cite{Zyla:2020zbs,ALICE:2021mgk}. Therefore, the non-prompt $\Ds$ $\pt$-differential production yield and $\RAA$ are compared with those of prompt $\Ds$ and non-prompt $\Dzero$ mesons, as well as with theoretical models based on beauty-quark transport in the QGP.

\section{Experimental apparatus and analysis technique}
\label{sec:analysis}

% apparatus and data sample
The $\Ds$-mesons were reconstructed from their hadronic decays with the ALICE central barrel detectors, which cover the full azimuth in the pseudorapidity interval $|\eta| < 0.9$ and are embedded in a large solenoidal magnet providing a uniform $0.5~\mathrm{T}$ magnetic field parallel to the beam direction.
Charged-particle trajectories are reconstructed from their hits in the Inner Tracking System (ITS)~\cite{ALICE:2010tia} and the Time Projection Chamber (TPC)~\cite{Alme:2010ke}.
Particle identification (PID) is provided via the measurement of the specific ionisation energy loss $\dEdx$ in the TPC and of the flight time of the particles from the interaction point to the Time-Of-Flight detector (TOF)~\cite{Akindinov:2013tea}.
The reconstruction of the interaction vertex and of the decay vertices of charm- and beauty-hadron decays relies on the precise determination of the track parameters in the vicinity of the interaction point provided by the ITS.

The data sample of Pb--Pb collisions used in the analysis was collected with the ALICE detector in 2018, during LHC Run 2. Three trigger classes were considered: minimum bias, central, and semicentral, all based on the signals in the two scintillator arrays of the V0 detector~\cite{ALICE:2013axi}, which covers the full azimuth in the pseudorapidity intervals $-3.7< \eta <-1.7$ (V0C) and $2.8< \eta <5.1$ (V0A). Background events due to the interaction of one of the beams with residual gas in the vacuum tube and other machine-induced backgrounds were rejected offline using the timing information provided by the V0 and the  neutron Zero Degree Calorimeters (ZDC)~\cite{Abelev:2014ffa}.
Only events with a primary vertex reconstructed within $\pm10$~cm from the centre of the detector along the beam-line direction were considered in the analysis.
Collisions were classified into centrality intervals, defined in terms of percentiles of the hadronic Pb--Pb cross section, based on the V0 signal amplitude as described in detail in Ref.~\cite{Adam:2015ptt}.
The measurement of non-prompt $\Ds$-meson production was carried out for central (0--10\%) and semicentral (30--50\%) collisions.
The number of events considered for the analysis is about $100 \times 10^6$ and $85 \times 10^6$ in the 0--10\% and 30--50\% centrality intervals, corresponding to integrated luminosities $\mathcal{L}_{\rm int}$ of $(130.5 \pm 0.5)~\upmu \mathrm{b}^{-1}$ and $(55.5 \pm 0.2)~\upmu \mathrm{b}^{-1}$, respectively~\cite{Loizides:2017ack}.
The average values of the nuclear overlap function, $\av{\TAA}$, for the considered central and semicentral event intervals were estimated via Glauber-model~\cite{dEnterria:2020dwq} simulations anchored to the V0 signal amplitude distribution, and are $(23.26\pm0.17)$ mb$^{-1}$ and $(3.92\pm0.06)$ mb$^{-1}$~\cite{ALICE-PUBLIC-2018-011, Loizides:2017ack}, respectively.

% analysis part
The $\Ds$ mesons and their charge conjugates were reconstructed via the $\DstophipitoKKpi$ decay channel  with branching ratio $\mathrm{BR} = (2.24 \pm 0.08) \%$~\cite{Zyla:2020zbs}. The analysis was based on the reconstruction of decay-vertex topologies displaced from the interaction vertex. For prompt mesons, the separation between the interaction point and the $\Ds{}$ decay vertex is governed by the mean proper decay length $c\tau$ of $\Ds{}$ mesons, which is about $151~\mum$~\cite{Zyla:2020zbs}. The decay vertices of non-prompt $\Ds{}$ mesons on average are more displaced than those of prompt $\Ds{}$ mesons due to the large mean proper decay lengths of beauty hadrons ($c\tau \simeq 450~\mum$~\cite{Zyla:2020zbs}). Therefore, by exploiting the selection of displaced decay-vertex topologies, it is possible to separate non-prompt $\Ds{}$ mesons from the combinatorial background and from prompt $\Ds{}$ mesons.

$\Ds$-meson candidates were built combining triplets of tracks with the proper charge signs, each with $|\eta| < 0.8$, at least 70 (out of a maximum of 159) crossed TPC pad rows, a track fit quality $\chi^{2}/{\rm ndf} < 1.25$ in the TPC (where ndf is the number of degrees of freedom involved in the track fit procedure), and a minimum of two (out of a maximum of six) hits in the ITS, with at least one in either of the two innermost layers, which provide the best pointing resolution. Moreover, at least 50 clusters
available for particle identification in the TPC were required, and only tracks with $\pt$ above $0.6~(0.4)~\gev/c$ were considered for central (semicentral) collisions. 
These track selection criteria limit the $\Ds$-meson acceptance in rapidity, which drops steeply to zero for $|y|>0.5$ at low $\pt$ and for $|y|>0.8$ at $\pt>5~\gevc$. Thus, only $\Ds$-meson candidates within a $\pt{}$-dependent fiducial acceptance region, $|y| < y_{\rm fid}(\pt)$, were selected. The $y_{\mathrm{fid}}(\pt)$ value was defined as a second-order polynomial function, increasing from 0.5 to 0.8 in the transverse-momentum range $0 < \pt < 5~\gevc$, and as a constant term, $y_{\mathrm{fid}}=0.8$, for $\pt > 5~\gevc$.

Similarly to other recent D-meson measurements by the ALICE Collaboration~\cite{ALICE:2021mgk, ALICE:2022tji, ALICE:2021kfc}, Boosted Decision Trees (BDT) algorithms were employed to reduce the large combinatorial background and to separate the contribution of prompt and non-prompt $\Ds$ mesons through a multiclass classification. In particular, the implementation of the BDT algorithm provided by the XGBoost~\cite{Chen:2016XST,barioglio_luca_2021_5070132} library was used. 
Background samples for the BDT training were extracted from the sidebands of the candidate invariant mass distributions in the data, namely from the $1.72 < M(\mathrm{KK}\uppi) < 1.83~\gev/c^2$ and $2.01 < M(\mathrm{KK}\uppi) < 2.12~\gev/c^2$ regions. Applying these selections, candidates belonging to $\DplustoKKpi$ decays are rejected.
Signal samples of prompt and non-prompt $\Ds$ mesons were obtained from Monte Carlo (MC) simulations. The MC samples were built by simulating Pb--Pb collisions with the HIJING~1.36~\cite{PhysRevD.44.3501} event generator in order to describe the charged-particle multiplicity and detector occupancy.
To enrich the sample of prompt and non-prompt D-meson signals, additional $\ccbar$- and $\bbbar$-quark pairs were injected into each HIJING event using the PYTHIA~8.243 event generator~\cite{Sjostrand:2006za, Sjostrand:2014zea} with Monash tune~\cite{Skands:2014pea}. The $\Ds{}$ mesons were forced to decay into the hadronic channel of interest for the analysis. The generated particles were then propagated through the apparatus using the GEANT3 transport code~\cite{Brun:1994aa}.
Detailed descriptions of the detector response, the geometry of the apparatus and the conditions of the luminous region, including their evolution with time during the data taking period, were included in the simulation.
Before the BDT training, loose kinematic and topological selections were applied to the $\Ds$-meson candidates together with the particle identification of decay-product tracks.
The $\Ds$-meson candidate information provided to the BDTs, as an input for the models to distinguish among prompt and non-prompt mesons and background candidates, was mainly based on the displacement of the tracks from the primary vertex, the distance between the $\Ds$-meson decay vertex and the primary vertex, the $\Ds$-meson impact parameter, and the cosine of the pointing angle between the $\Ds$-meson candidate line of flight (the vector connecting the primary and secondary vertices) and its reconstructed momentum vector. In addition, the absolute difference between the reconstructed ${\rm K^+K^-}$ invariant mass and the PDG average mass for the $\upphi$ meson~\cite{Zyla:2020zbs} and variables related to the PID of decay tracks were also included.
Independent BDTs were trained in the different $\pt$ intervals of the analysis and for the different centrality intervals. Subsequently, they were applied to the real data sample in which the type of candidate is unknown. The BDT outputs are related to the candidate probability to be a non-prompt $\Ds$ meson or combinatorial background. Selections on the BDT outputs were optimised to obtain a high non-prompt $\Ds$-meson fraction while maintaining a reliable signal extraction from the candidate invariant mass distributions.

\begin{figure}[tb]
  \begin{center}
  \includegraphics[width = 0.5\textwidth]{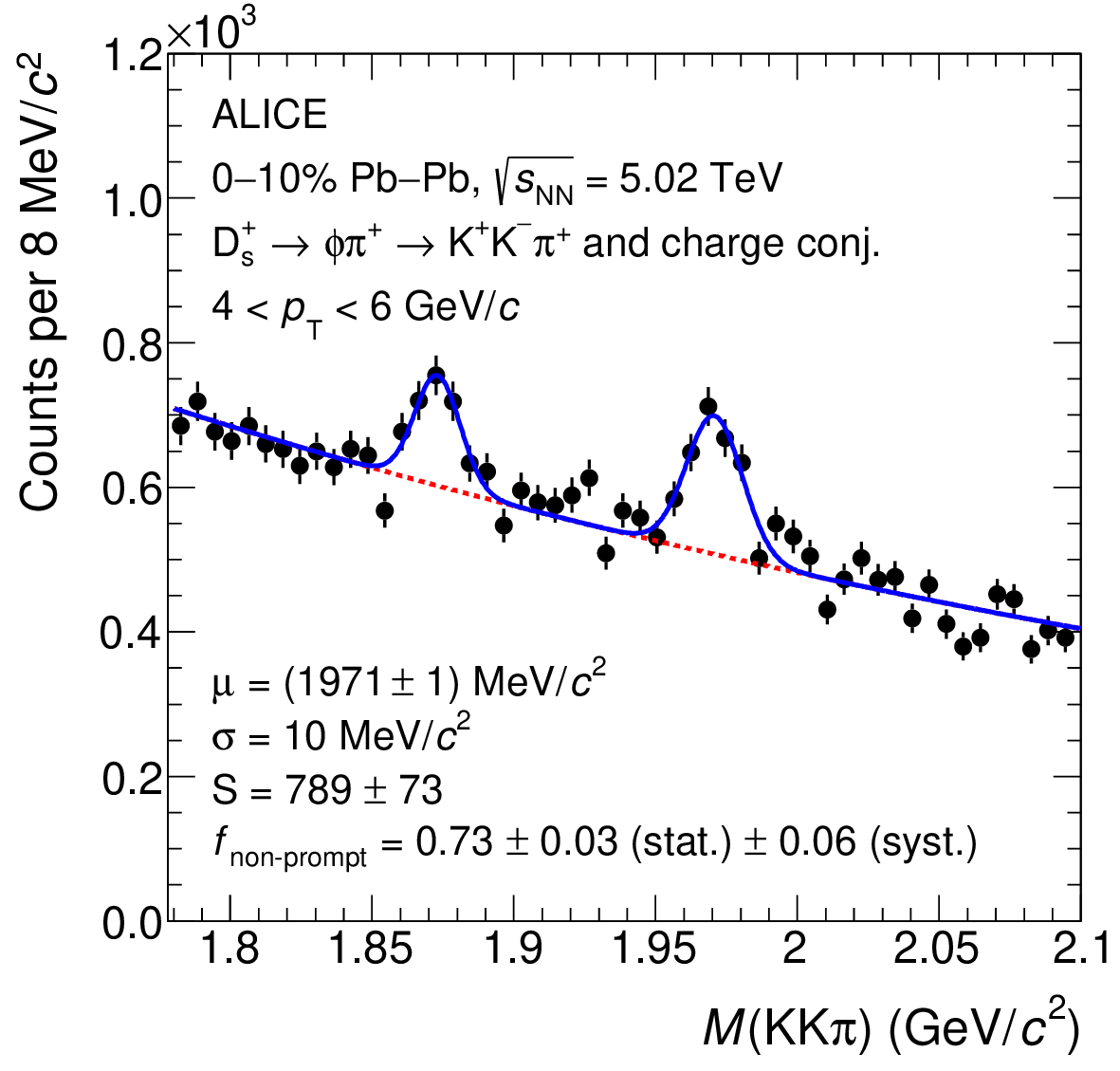}
  \end{center}
  \caption{Invariant mass distribution of non-prompt $\Ds$ candidates and their charge conjugates in the $4<\pt<6~\gevc$ interval for central Pb--Pb collisions. The blue solid line shows the total fit function and the red dashed line the combinatorial-background contribution. The values of the mean ($\upmu$), width ($\sigma$), and raw yield ($S$) of the signal peak are reported together with their statistical uncertainties resulting from the fit. The fraction of non-prompt candidates in the measured raw yield is reported with its statistical and systematic uncertainties.}
  \label{fig:inv_mass}
\end{figure}

The $\Ds$-meson candidates were selected by requiring a high probability to be non-prompt $\Ds$ mesons and a low probability to be combinatorial background. 
The raw yield of $\Ds$ mesons, including both particles and antiparticles, was extracted from binned maximum-likelihood fits to the invariant mass ($M$) distributions in transverse-momentum intervals $4<\pt<36~\gev/c$ and $2<\pt<24~\gev/c$ for the 0--10\% and the 30--50\% centrality intervals, respectively. The fit function was composed of a Gaussian for the description of the signal and an exponential term for the background. An additional Gaussian was used to describe the peak due to the decay $\DplustoKKpi$, with a branching ratio of $(9.68\pm 0.18) \times 10^{-3}$~\cite{Zyla:2020zbs}, present at a lower invariant mass value than the $\Ds$-meson signal peak.
To improve the stability of the fits, the width of the $\Ds$-meson signal peak was fixed to the value extracted from a data sample dominated by prompt candidates, which is characterised by a signal extraction with higher statistical significance. 
As an example, the invariant mass distribution for the $4<\pt<6~\gevc$ interval in central Pb--Pb collisions, together with the result of the fit and the estimated non-prompt fraction is reported in Fig.~\ref{fig:inv_mass}. 
The measured raw yield, although dominated by non-prompt candidates, still contains a residual contribution of prompt $\Ds$ mesons which satisfy the BDT-based selections. The procedure used to calculate the fraction of non-prompt candidates present in the extracted raw yield is described below.
The statistical significance of the observed signals varies from about 4 to 11 depending upon the $\pt$ and centrality intervals. 

The corrected $\pt$-differential yields of non-prompt $\Ds$ mesons were computed for each $\pt$ interval as
\begin{equation}
  \label{eq:dNdpt}
  \left.\frac{{\rm d} N}{{\rm d}\pt}\right|_{|y|<0.5}= \frac{1}{2}\times
  \frac{1}{\Delta \pt}\times\frac{\left.\fnonprompt(\pt)\times N^{\rm D+\overline D,raw}(\pt)\right|_{|y|<y_{\rm fid}(\pt)}}{c_{\Delta y}(\pt)\times\effNP{}(\pt)\times\mathrm{BR}\times N_\mathrm{evt}}\,.
\end{equation}
The raw-yield values $N^{\rm D+\overline D,raw}$ were divided by a factor of two and multiplied by the non-prompt fraction $\fnonprompt$ to obtain the charge-averaged yields of non-prompt $\Ds$ mesons. Furthermore, they were divided by the acceptance-times-efficiency correction factor of non-prompt $\Ds$ mesons $\effNP{}$, the BR of the decay channel, the width of the $\pt$ interval $\Delta \pt$, the correction factor for the rapidity coverage $c_{\Delta y}$, and the number of analysed events $N_{\rm evt}$.
The correction factor for the rapidity acceptance $c_{\Delta y}$ was defined as the ratio between the generated D-meson yield in $\Delta y = 2\,y_{\rm fid}(\pt{})$ and that in $|y|<0.5$. It was computed with FONLL perturbative QCD calculations~\cite{Cacciari:1998it,Cacciari:2001td} as in Refs.~\cite{ALICE:2021kfc, ALICE:2021rxa}.

\begin{figure}[tb]
  \begin{center}
  \includegraphics[width = 0.5\textwidth]{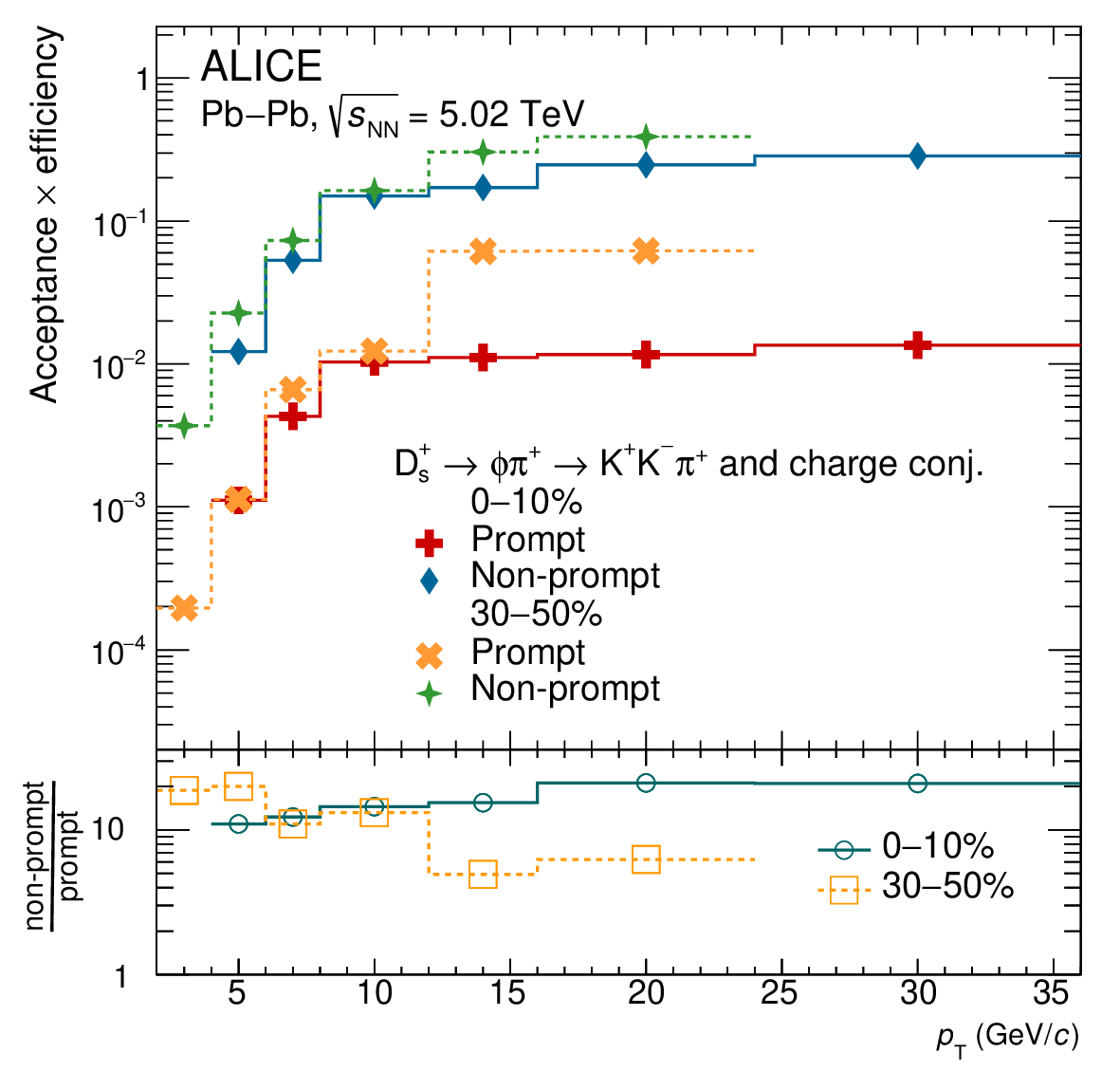}
  \end{center}
  \caption{Acceptance-times-efficiency factors for prompt and non-prompt $\Ds$ mesons as a function of $\pt$ in the 0--10\% and 30--50\% centrality intervals, together with their ratios (bottom panel).}
  \label{fig:eff}
\end{figure}

The \AccEff correction factor was obtained from MC simulations, using samples not employed in the BDT training.
The $\Ds{}$-meson $\pt{}$ distributions from simulations were reweighed in order to mimic the realistic shapes in the determination of the \AccEff factor, which depends on $\pt{}$. In particular, weights were applied to the $\pt$ distributions of prompt $\Ds$ mesons and of beauty-hadron mother particles in case of non-prompt $\Ds$ mesons. These weights were defined to reproduce the shapes given by FONLL calculations multiplied by the $\RAA{}$ of prompt $\Ds$ mesons and B mesons predicted by the TAMU~\cite{He:2014cla,He:2019vgs} model. The TAMU model implements the charm- and beauty-quark transport inside a strangeness-rich QGP, and it reasonably reproduces the prompt D-meson measurements at low $\pt$~\cite{ALICE:2021kfc, ALICE:2021rxa}. 
The \AccEff factors as a function of $\pt$ for prompt and non-prompt $\Ds$ mesons in the 0--10\% and 30--50\% centrality intervals are displayed in Fig.~\ref{fig:eff}, along with the ratios of the non-prompt to prompt factors. The prompt $\Ds$-meson acceptance times efficiency is smaller than that of non-prompt $\Ds$ mesons by a factor varying from 5 to 20 depending on $\pt$ and centrality. This is expected since the selections applied to obtain the non-prompt enriched sample strongly suppress the prompt $\Ds$-meson efficiency. Instead, the acceptance is the same for prompt and non-prompt mesons. In central collisions, the prompt $\Ds$-meson suppression increases with increasing $\pt$. The opposite trend is observed in semicentral collisions, since less stringent selections on the BDT outputs are necessary to extract the non-prompt $\Ds$-meson signal due to the lower yield.

\begin{figure}[tb]
      \begin{center}
      \includegraphics[width = 0.5\textwidth]{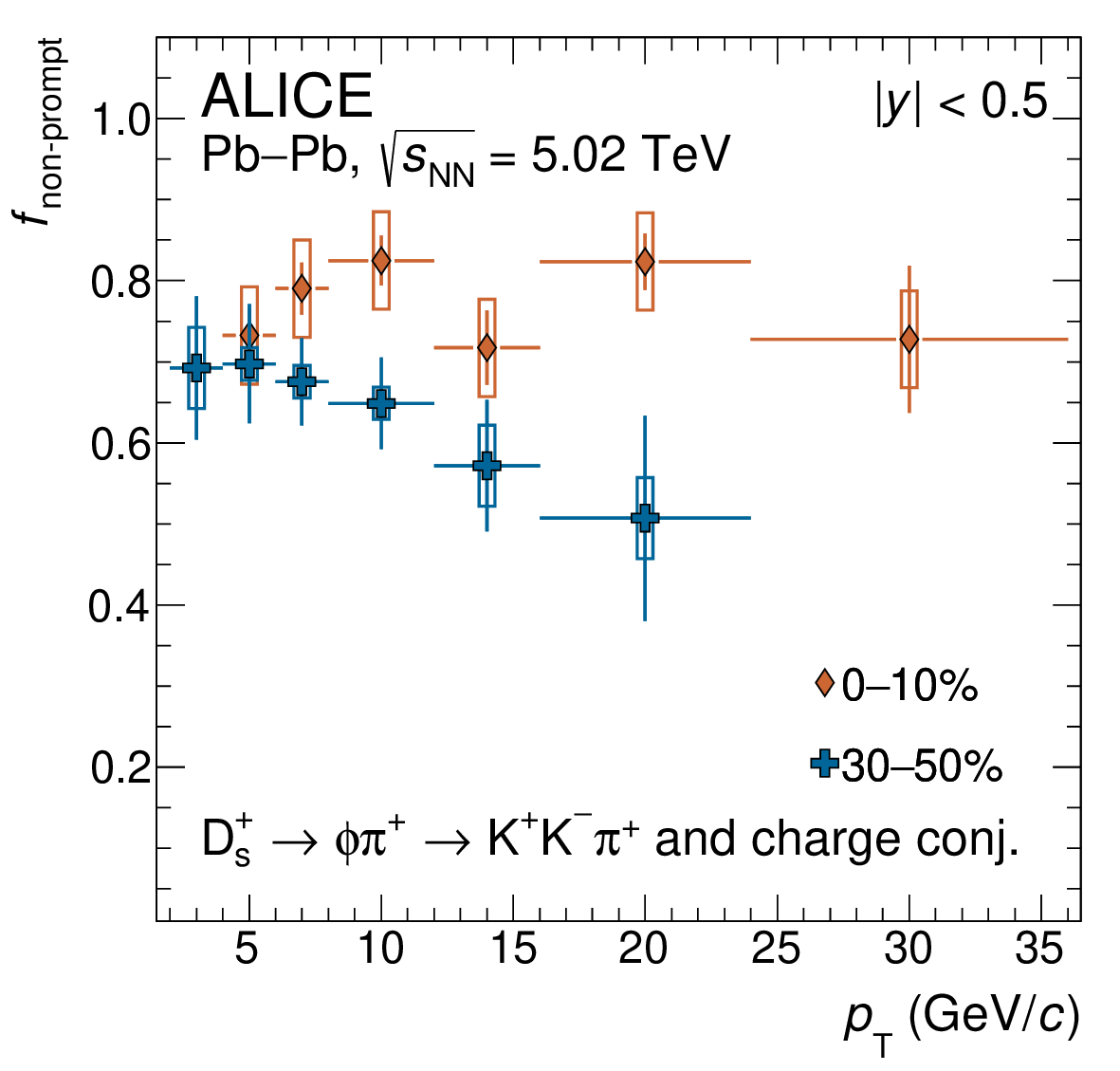}
      \end{center}
      \caption{Fraction of non-prompt $\Ds$ mesons in the extracted raw yield as a function of $\pt$ in the 0--10\% and 30--50\% centrality intervals. The vertical bars (boxes) report the statistical (systematic) uncertainties.}
      \label{fig:non_prompt_frac}
    \end{figure}

The fraction $\fnonprompt$ of non-prompt $\Ds$ mesons in the extracted raw yield was estimated with a data-driven procedure based on the construction of data samples with different abundances of prompt and non-prompt candidates. These samples were built by varying the selection on the BDT output related to the candidate probability to be a non-prompt $\Ds$ meson. Starting from the values of raw yield and acceptance times efficiency of prompt and non-prompt $\Ds{}$ mesons obtained for each sample, the corrected yield of prompt and non-prompt $\Ds{}$ mesons and the $\fnonprompt$ fraction were calculated. This data-driven technique does not depend on theoretical calculations of heavy-quark production and interaction with the QGP constituents, and it is described in detail in Ref.~\cite{ALICE:2021mgk}.
The $\fnonprompt$ fractions obtained as a function of $\pt$ in central and semicentral Pb--Pb collisions are reported in Fig.~\ref{fig:non_prompt_frac}, together with their statistical and systematic uncertainties. The determination of the systematic uncertainty on the $\fnonprompt$ fraction is described in Section~\ref{sec:syst}. The $\fnonprompt$ values vary between about 0.72 (0.56) and 0.82 (0.70) in the 0--10\% (30--50\%) centrality interval as a function of transverse momentum. The $\fnonprompt$ is observed to be on average lower in semicentral collision with respect to central collisions. This difference is expected as in the 30--50\% centrality interval less stringent BDT selections were applied compared to 0--10\% centrality interval. 

The non-prompt $\Ds{}$-meson nuclear modification factor, $\RAA{}$, was computed according to Eq.~\ref{eq:RAA}. The measurement of the $\pt{}$-differential cross section of non-prompt $\Ds{}$ mesons at midrapidity ($|y| < 0.5$) in pp collisions at $\sqrt{s} = 5.02~\mathrm{TeV}$ from Ref.~\cite{ALICE:2021mgk}, which covers the transverse-momentum interval $2 < \pt{} < 12~\gevc{}$, was used as the reference for the $\RAA{}$ computation. 
For $\pt>12~\gevc{}$, an extrapolated pp reference was obtained from FONLL calculations of the beauty-hadron cross section and by using PYTHIA~8 to describe the decay kinematics of  beauty hadrons to $\Ds$ mesons, for more details see Ref.~\cite{ALICE:2021mgk}. The resulting predictions were then scaled to match the measured values at lower transverse momenta. The total systematic uncertainty on the pp reference is $^{+38}_{-28}\%$ for all the extrapolated $\pt{}$ intervals. The procedures for the $\pt{}$ extrapolation and the systematic uncertainty estimation are the same as in Ref.~\cite{Adam:2015sza}.

\section{Systematic uncertainties}
\label{sec:syst}
The following sources of systematic uncertainty were considered for the production yield and $\RAA$ estimation: (i) the raw-yield extraction, (ii) track reconstruction efficiency, (iii) non-prompt $\Ds$-meson fraction, (iv) BDT selection efficiency, (v) PID selection efficiency, (vi) relative abundances of beauty-hadron species in the MC simulation, and (vii) shapes of the simulated $\pt$-differential distributions. The resulting systematic uncertainties on the non-prompt $\Ds$-meson yield and $\RAA$ in representative $\pt$ intervals are summarised in Table~\ref{tab:systDs}.
In the $\RAA$ computation, the systematic uncertainties on the pp measurement were treated as uncorrelated from the ones on the Pb–Pb corrected yields, except for the uncertainty on the BR (3.6\%)~\cite{Zyla:2020zbs} which cancels in the $\RAA$ and was considered only in the $\pt$-differential production yield. The normalisation uncertainty on the $\RAA$ includes the uncertainty on the integrated luminosity in pp collisions (2.1\%~\cite{ALICE-PUBLIC-2018-014}), the uncertainty on the $\langle\TAA\rangle$ estimation, 0.7\% (1.5\%) for the 0–10\% (30–50\%) centrality interval~\cite{ALICE-PUBLIC-2018-011}, and the one related to the
centrality-interval definition. This last contribution is due to the uncertainty on the fraction of the hadronic cross section used in the Glauber fit to determine the centrality. It was estimated to be < 0.1\% and
2\% for the 0–10\% and 30–50\% centrality intervals, respectively~\cite{Adam:2015sza}.

The systematic uncertainty on the raw-yield extraction was estimated by adopting several fit configurations changing the background fit function (linear and parabolic), the upper and lower fit limits, and the bin size of the invariant mass spectrum. The sensitivity to the line shape of the $\Ds$ peak was tested by comparing the raw-yield values from the fits with those obtained by counting the candidates in the invariant mass region of the signal after subtracting the background estimated from the side bands.

The systematic uncertainty on the track reconstruction efficiency accounts for possible discrepancies between data and MC in the ITS--TPC prolongation efficiency and in the selection efficiency due to track-quality criteria in the TPC. The per-track systematic uncertainties were estimated by varying the track-quality selection criteria and by comparing the prolongation probability of the TPC tracks to the ITS hits in data and simulations. They were then propagated to the non-prompt $\Ds$ mesons via their decay kinematics.

The systematic uncertainties on the non-prompt $\Ds$-meson fraction and the BDT selection efficiency are due to possible discrepancies between data and MC in the distributions of the variables used in the BDT-model training (i.e.~the $\Ds$-meson decay-vertex topology, kinematic, and PID variables). The former was computed by varying the configuration and the number of BDT selections employed in the data-driven method described in Sec.~\ref{sec:analysis}. In particular, wider and narrower intervals of the probability to be non-prompt $\Ds$ mesons, and smaller and larger step sizes between the chosen BDT selections were considered. For each configuration, the non-prompt $\Ds$-meson fraction was recomputed. The systematic uncertainty related to the BDT selection efficiency was studied by repeating the entire analysis varying the selection criteria based on the BDT outputs. The uncertainty for this source of systematic uncertainty was assigned considering the RMS and the shift of the corrected yield obtained by varying the BDT selection with respect to the reference one.

Analogously, the systematic uncertainty on the PID selection efficiency relative to the loose selection on the PID variables applied before the BDT ones was also considered. This source was evaluated in the prompt $\Ds$-meson analysis~\cite{ALICE:2021kfc}, and it was found to be negligible for the adopted PID strategy.

The selection efficiency of non-prompt $\Ds$ mesons originating from the decay of different beauty-hadron species can differ because of the different lifetime of the parent hadron and the different decay kinematics. Consequently, an imperfect description in the MC simulation of the beauty-hadron composition might result in a bias in the estimation of the D-meson efficiencies. This is especially important for $\Ds$ mesons, which receive significant contributions from all the three ground-state B-meson species ($\Bplus$, $\Bzero$, and $\Bs$). The PYTHIA~8 event generator describes the measurements of different B-meson species in pp collisions~\cite{ALICE:2021mgk}, however in heavy-ion collisions an enhanced production of strange over non-strange B mesons is expected compared to the one observed in pp collisions. Nevertheless, since no precise measurement of $\Bs$-meson production down to low momentum is available in Pb--Pb collisions, the relative abundances present in PYTHIA~8 were used without applying any reweighting. The systematic uncertainty introduced by this assumption was estimated by reweighting the $\Bs$ contribution present in the MC enhanced by a factor 2 as predicted by the TAMU model~\cite{He:2014cla}. The systematic uncertainty was assigned considering the variation between the production yield estimated using the enhanced $\Bs$ contribution and the default one.

The systematic uncertainty due to the shape of the $\pt$ distributions of $\Ds$ mesons and beauty hadrons in the MC simulations was evaluated by applying different weights to the $\pt$ distributions of prompt $\Ds$ mesons and of beauty-hadron mother particles in case of non-prompt $\Ds$ mesons. As an alternative to the TAMU model, the shape resulting from the LIDO model~\cite{Ke:2018tsh} was considered. The main difference between the TAMU and LIDO model derives from the fact that the former includes the enhanced production of the $\Bs$ mesons, unlike the latter. An additional variation of the shape of the $\pt$ distributions of prompt $\Ds$ mesons was included considering the results from Ref.~\cite{ALICE:2021kfc}. 
The systematic uncertainty was assigned considering the variation of the corrected yield compared to the default case.

\begin{table}[!tb]
\caption{Systematic uncertainties on the measurement of the non-prompt $\Ds$-meson corrected yield and $\RAA$ in the 0--10\% and 30--50\% centrality intervals for representative transverse-momentum intervals.} 
\label{tab:systDs}
\centering
\begin{tabular}{l c c | c c} 
\toprule
Centrality interval & \multicolumn{2}{c}{0--10\%} & \multicolumn{2}{c}{30--50\%}\\
%\cmidrule(lr){2-3}   \cmidrule(lr){4-5}
$\pt~(\GeV/c)$ & 4--6 & 12--16 & 2--4 & 12--16\\
\midrule
Yield extraction & 5\% & 5\% & 10\% & 5\% \\

Tracking efficiency & 13\% & 13\% & 11\% & 12\% \\

Non-prompt fraction & 6\% & 6\% & 5\% & 6\% \\

Selection efficiency & 8\% & 5\% & 10\% & 5\% \\

PID efficiency & negl. & negl. & negl. & negl. \\

B hadrochemistry & 1\% & 1\% & 1\% & 1\%\\

MC $\pt$ shape & 10\% & 8\% & 15\% & 2\% \\

\midrule

Centrality limits & \multicolumn{2}{c}{ < 0.1\%} & \multicolumn{2}{c}{2\%}\\

$\langle \TAA \rangle$ & \multicolumn{2}{c}{0.7\%} & \multicolumn{2}{c}{1.5\%}\\

$\mathcal{L}_\mathrm{int}^\mathrm{pp}$ & \multicolumn{4}{c}{2.1\%} \\

Branching ratio & \multicolumn{4}{c}{3.6\%} \\
\bottomrule
\end{tabular}
\end{table}

\section{Results}
\label{sec:results}
\begin{figure}[!tb]
\begin{center}
    \includegraphics[width=0.5\textwidth]{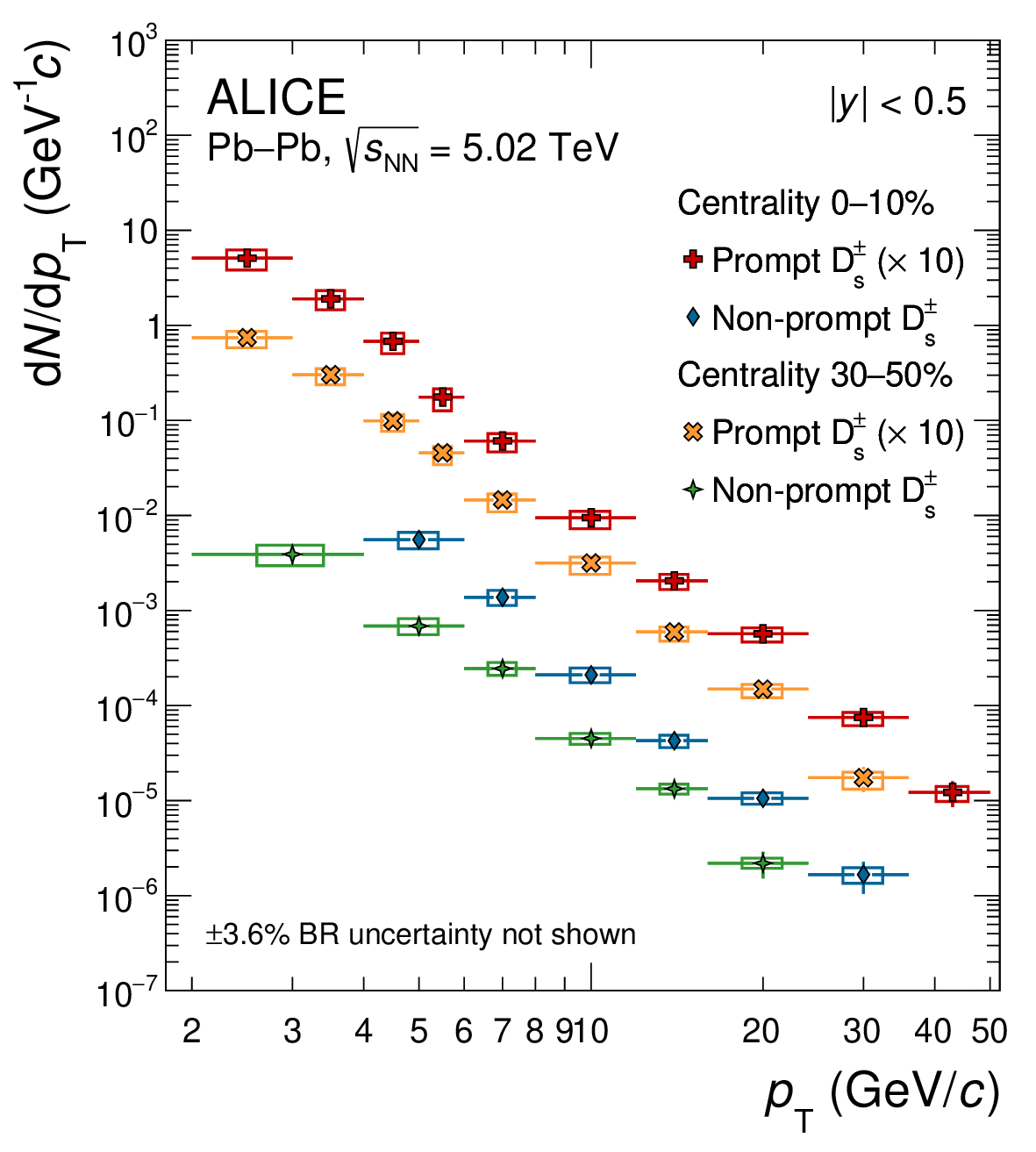}
    \caption{Prompt and non-prompt $\Ds$ meson production yield in central and semicentral Pb--Pb collisions at $\sqrtsNN = 5.02~\TeV$. The prompt $\Ds$ results are taken from Ref.~\cite{ALICE:2021kfc} and scaled by a factor 10 for visibility. The vertical bars (boxes) report the statistical (systematic) uncertainties.}
    \label{fig:prompt_nonprompt_Ds_corrYields}
\end{center}
\end{figure}

Figure~\ref{fig:prompt_nonprompt_Ds_corrYields} shows the $\pt$-differential production yield of prompt and non-prompt $\Ds$ mesons in central and semicentral Pb--Pb collisions at $\sqrtsNN = 5.02~\TeV$. The measured prompt $\Ds$-meson production yields were taken from Ref.~\cite{ALICE:2021kfc} and scaled by a factor 10 for visibility.

\begin{figure}[!tb]
\begin{center}
    \includegraphics[width=0.9\textwidth]{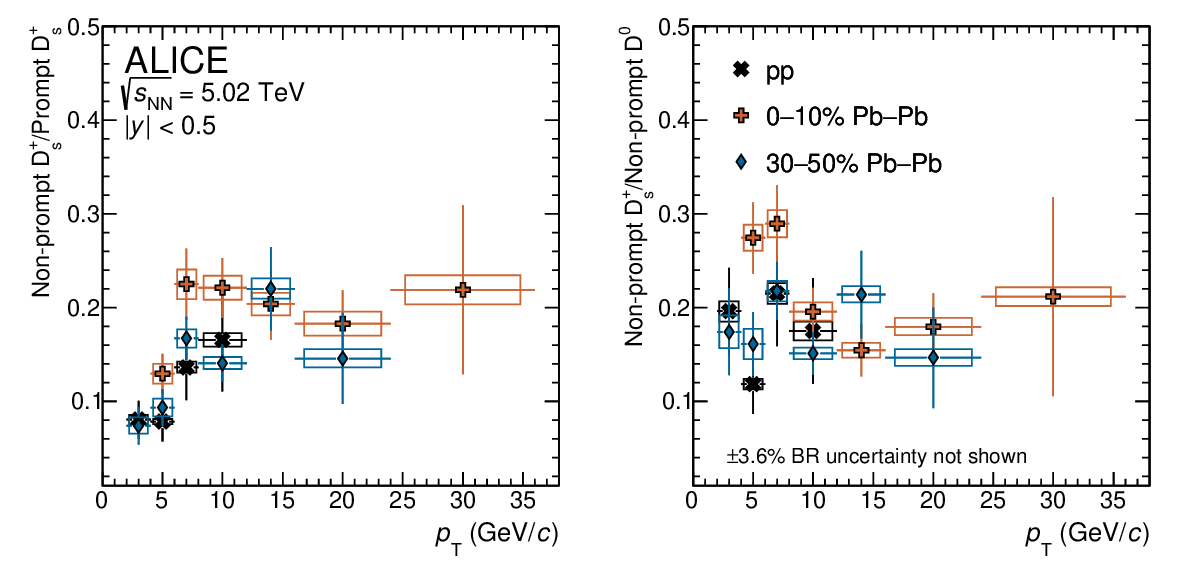}
    \caption{The $\pt$-differential production yield of non-prompt $\Ds$ mesons divided by those of prompt $\Ds$ mesons (left panel) and non-prompt $\Dzero$ mesons (right panel) for the 0--10\% and 30--50\% centrality intervals in Pb--Pb collisions at $\sqrtsNN=5.02~\TeV$ from Refs.~\cite{ALICE:2021kfc, ALICE:2022tji} compared with those in pp collisions at the same centre-of-mass energy from Ref.~\cite{ALICE:2021mgk}.}
    \label{fig:CorrYratio}
\end{center}
\end{figure}

Figure~\ref{fig:CorrYratio} reports the ratios of the production yield of non-prompt to prompt $\Ds$ (left panel) and non-prompt $\Ds$ to non-prompt $\Dzero$~\cite{ALICE:2022tji} (right panel) in central and semicentral Pb--Pb collisions, as well as in pp collisions~\cite{ALICE:2021mgk}. Computing these ratios helps to further investigate the effects of the QGP medium on the hadron formation mechanism. To get an indication of the $\Bs$-meson $\pt$ probed by non-prompt $\Ds$ mesons, a simulation with PYTHIA~8 was performed. As an example, the mean $\pt$ distribution of $\Bs$ mesons decaying to $\Ds$ mesons with $4<\pt<6~\GeV/c$ has a mean of about $8.8~\GeV/c$ and an RMS of about $3.1~\GeV/c$.
The non-prompt to prompt $\Ds$-meson ratio ranges between about 0.05 and 0.20 and increases with increasing $\pt$ up to $\pt = 10~\GeV/c$. At higher momentum the slope of the ratios seems to reduce, even though no firm conclusions can be drawn with the current uncertainties. On the other hand, the non-prompt $\Ds$ to non-prompt $\Dzero$ ratio shows an almost flat trend around 0.2 in the $\pt$ range of the measurement. The ratios computed in pp and semicentral Pb--Pb collisions are compatible within the uncertainties. A hint of enhancement compared to pp collisions with a significance of 1.7$\sigma$, where $\sigma$ indicates the sum in quadrature of statistical and systematic uncertainties, is found by performing a weighted average of the non-prompt $\Ds/\Dzero$ values in the $4 < \pt < 12~\GeV/c$ interval for the 0--10\% centrality class. The inverse of the squared sum of the relative statistical and $\pt$-uncorrelated systematic uncertainties was used as weight in the average. All the systematic uncertainties, except for those on the raw-yield extraction, were considered as fully correlated in $\pt$. This hint of a larger non-prompt $\Ds/\Dzero$ yield ratio is consistent with an enhanced production of strange-beauty mesons in heavy-ion collisions compared to pp collisions, as expected in a scenario in which beauty quarks hadronise via recombination with surrounding quarks in the strangeness-enriched QGP medium. In the transverse-momentum interval $4<\pt<12~\GeV/c$, also the non-prompt to prompt $\Ds$-meson ratio in the 0--10\% centrality class shows a mild enhancement with respect to pp collisions with a significance of 1.6$\sigma$.

\begin{figure}[!tb]
\begin{center}
    \includegraphics[width=0.75\textwidth]{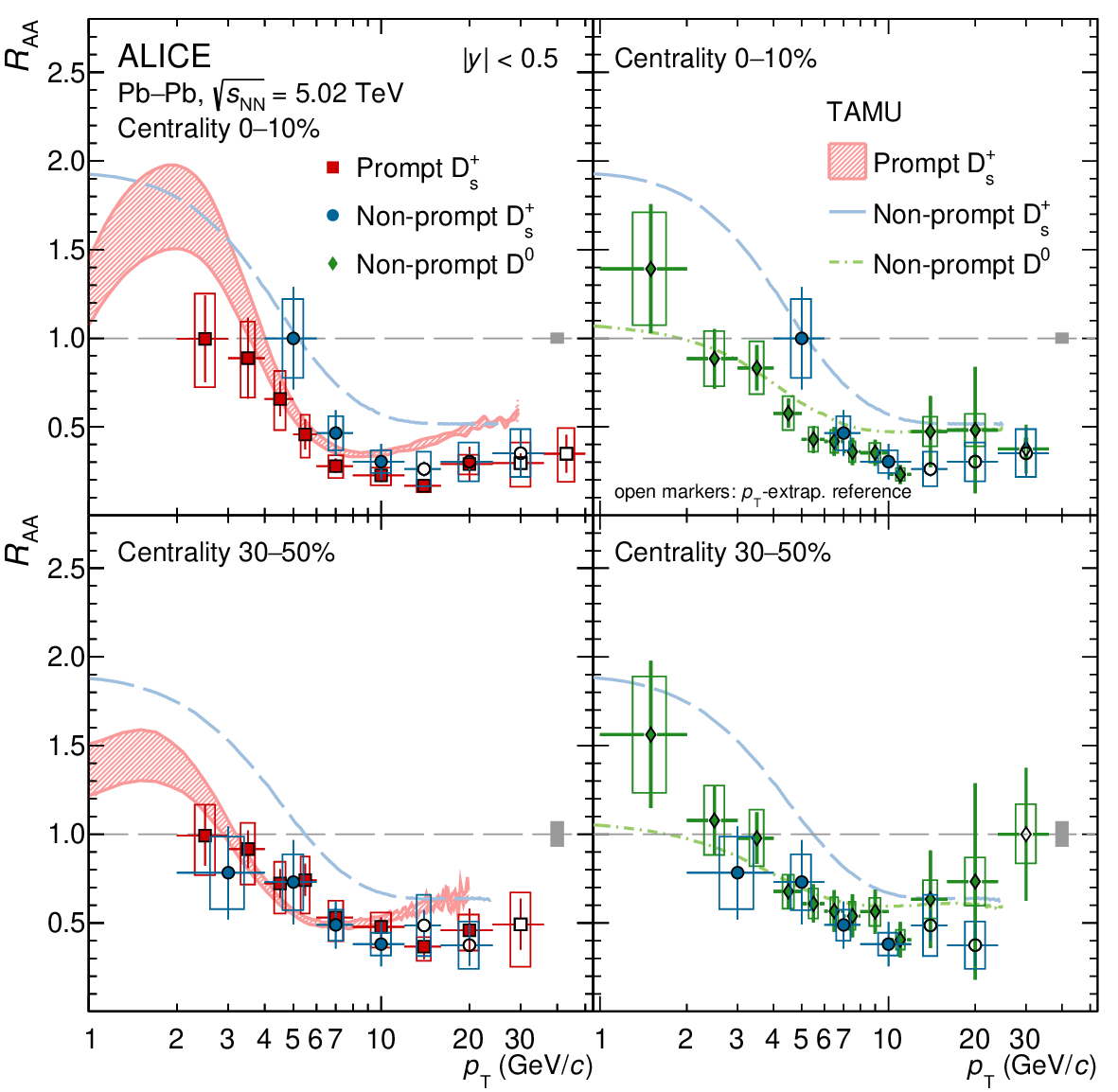}
    \caption{Left panels: prompt (Ref.~\cite{ALICE:2021kfc}) and non-prompt $\Ds$-meson $\RAA$ in central (top) and semicentral (bottom) Pb--Pb collisions at $\sqrtsNN = 5.02$ TeV. Right panels: non-prompt $\Ds$- and $\Dzero$-meson (Ref.~\cite{ALICE:2022tji}) $\RAA$ in central (top) and semicentral (bottom) Pb--Pb collisions at $\sqrtsNN = 5.02$ TeV. The experimental results are compared with the predictions of the TAMU model~\cite{He:2014cla}. Statistical (bars), systematic (boxes), and normalisation (shaded box around unity) uncertainties are shown.}
    \label{fig:RAARatios4panel}
\end{center}
\end{figure}

The $\RAA$ of non-prompt $\Ds$ mesons was computed according to Eq.~\ref{eq:RAA}, where the pp reference was obtained from the measurement published in Ref.~\cite{ALICE:2021mgk}.
To study the effects of the QGP medium on the resulting momentum spectra and the hadronisation mechanism of beauty quarks, the nuclear modification factor measured for the non-prompt $\Ds$ mesons was compared to that of prompt $\Ds$~\cite{ALICE:2021kfc} and non-prompt $\Dzero$~\cite{ALICE:2022tji} mesons measured at the same centre-of-mass energy per nucleon pair. The prompt and non-prompt $\Ds$ $\RAA$ are compared in the top- and bottom-left panels of Fig.~\ref{fig:RAARatios4panel} for the 0--10\% and 30--50\% centrality class, respectively. Analogously, the comparison between the nuclear modification factor of non-prompt $\Ds$ and non-prompt $\Dzero$ mesons is reported in the right panels of the same figure. The $\RAA$ of prompt and non-prompt D mesons shows a decreasing trend with increasing $\pt$ up to a minimum of about 0.2 (0.4) around $10~\GeV/c$ in the 0--10\% (30--50\%) centrality class. In the lowest $\pt$ intervals, the $\RAA$ increases up to unity. In particular, the central values of the non-prompt $\Ds$ $\RAA$ are higher with respect to those of prompt $\Ds$ and non-prompt $\Dzero$ in the 0--10\% centrality class for $\pt < 6~\GeV/c$, even though they are compatible within uncertainties. This possible difference between prompt and non-prompt $\Ds$ $\RAA$ would be consistent with the different loss of energy experienced by charm and beauty quarks traversing the QGP. In fact, the effect due to the different decay kinematics of charm and beauty hadrons is found to be negligible, as discussed in Ref.~\cite{ALICE:2022tji}. 
Instead, the difference between non-prompt $\Ds$ and $\Dzero$ mesons could result from the hadronisation via recombination and the presence of a strangeness-rich environment. In semicentral collisions, no separation among the $\RAA$ of prompt $\Ds$, non-prompt $\Ds$, and non-prompt $\Dzero$ is observed within the measurement uncertainties.

The $\RAA$ measurements were compared with the predictions of the TAMU model~\cite{He:2014cla}. In the TAMU model, the heavy-quark transport is described via the Langevin equation and the hadronisation can occur both via recombination with light quarks from the medium, which is the dominant mechanism at low $\pt$, or via fragmentation, which becomes more important at high $\pt$. The TAMU predictions are shown in Fig.~\ref{fig:RAARatios4panel}. The uncertainty band for prompt $\Ds$ mesons is due to the modification of the parton distribution functions in Pb nuclei, which is neglected for the beauty-quark production. The TAMU model qualitatively describes the $\pt$ trend of the non-prompt $\Ds$-meson $\RAA$, although it overestimates the measurements.

\begin{figure}[!tb]
\begin{center}
    \includegraphics[width=0.9\textwidth]{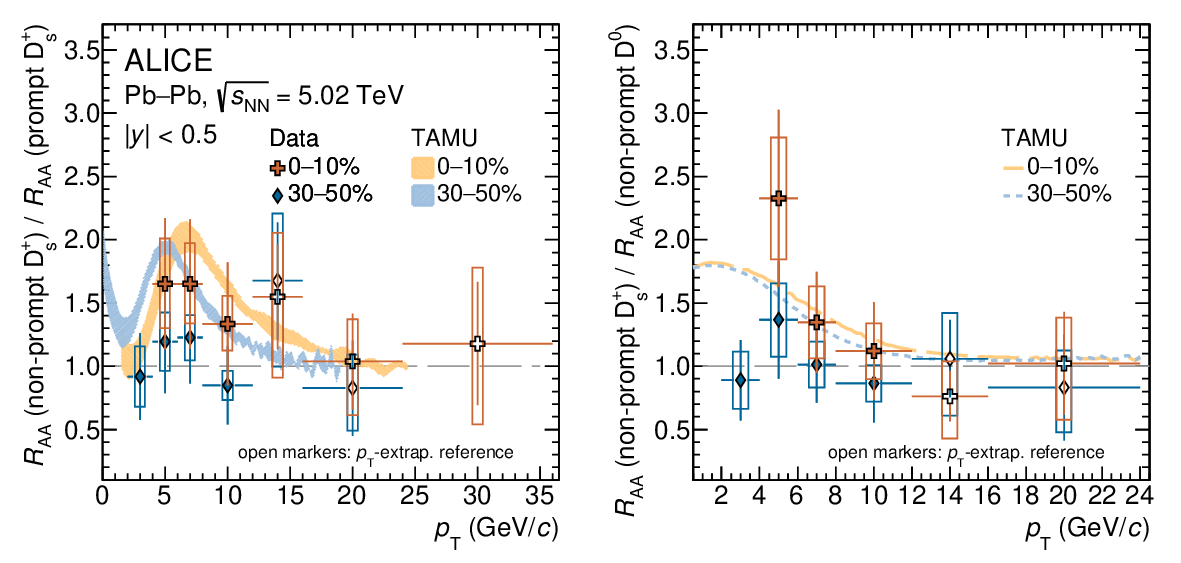}%
    \caption{The $\RAA$ of non-prompt $\Ds$ mesons divided by the one of prompt $\Ds$ mesons~\cite{ALICE:2021kfc} (left panel) and non-prompt $\Dzero$ mesons~\cite{ALICE:2022tji} (right panel) for the 0--10\% and 30--50\% centrality intervals in Pb--Pb collisions at $\sqrtsNN=5.02~\TeV$. The measurements are compared with TAMU model predictions~\cite{He:2014cla}. Statistical (bars) and systematic (boxes) uncertainties are shown.}
    \label{fig:RAARatio}
\end{center}
\end{figure}

In the left and right panels of Fig.~\ref{fig:RAARatio}, the nuclear modification factors of non-prompt $\Ds$ mesons divided by that of prompt $\Ds$ mesons and non-prompt $\Dzero$ mesons are shown, respectively. The measurements in both centrality intervals are compared with the predictions of the TAMU model. In the 0--10\% centrality class, the non-prompt $\Ds$ to prompt $\Ds$ $\RAA$ ratio suggests a hint of enhancement with a statistical significance of 1.6$\sigma$ in the  $4 < \pt < 12~\GeV/c$ interval, which is by construction the same of that reported for the corresponding yield ratio. The $\RAA$ ratio is consistent with a larger energy loss for the charm quark with respect to the beauty quark due to its smaller mass, as already suggested by the results shown in Fig.~\ref{fig:RAARatios4panel}. No hint for a ratio of the $\RAA$ larger than unity is observed in semicentral collisions. Considering the measurement uncertainties, TAMU predictions qualitatively describe the results for central collisions. At variance, for semicentral collisions the TAMU model overestimates the $\RAA$ ratio values. The measurements of the non-prompt $\Ds$ to non-prompt $\Dzero$ $\RAA$ ratio suggest a possible enhancement with respect to unity in the $4 < \pt < 12~\GeV/c$ interval for central collisions, as reported for the yield ratio. In this case, the rise at low $\pt$ might be a consequence of the abundance of strange quarks thermally produced in the QGP and the dominance of the hadronisation via recombination in this range of momentum. The TAMU model describes the data within the experimental uncertainties.

\section{Conclusions}
\label{sec:conclusions}
In this Letter, the first measurement of the non-prompt $\Ds$-meson production at midrapidity in Pb--Pb collisions at $\sqrtsNN = 5.02$ TeV was reported.

The non-prompt $\Ds$-meson production yield was measured between 4 and 36 (2 and 24) $\GeV/c$ in the 0--10\% (30--50\%) centrality interval. These measurements were compared to the ones performed for prompt $\Ds$ and non-prompt $\Dzero$ mesons at the same centre-of-mass energy. The production yield was employed to compute the non-prompt $\Ds$-meson $\RAA$, which was compared with the $\RAA$ of prompt $\Ds$ and non-prompt $\Dzero$ mesons.

The non-prompt $\Ds$ $\RAA$ shows a significant $\pt$ dependence. A minimum at intermediate transverse momentum ($\pt\approx10~\GeV/c$) around 0.2 (0.4) in central (semicentral) collisions, and a mild increase with decreasing $\pt$, with $\RAA$ reaching (close to) unity at $\pt\approx4\text{--}6~ (2\text{--}4)~\GeV/c$ in the 0--10\% (30--50\%) centrality interval are reported. The TAMU model, which implements the parton in-medium energy loss through collisional processes as well as the beauty-quark hadronisation both via fragmentation and recombination, describes the $\pt$ trend of the $\RAA$. However, it overestimates the measurements. Further comparisons were performed between prompt and non-prompt $\Ds$ as well as non-prompt $\Dzero$ mesons by computing the ratios of their production yields and $\RAA$. These ratios suggest the presence of an enhancement of non-prompt $\Ds$ mesons compared to prompt $\Ds$ (non-prompt $\Dzero$) mesons in central collisions in the $4 < \pt < 12~\GeV/c$ interval, with a significance of 1.6$\sigma$ (1.7$\sigma$). The increase is consistent with expectations for the overall effect of the energy-loss mechanism and the hadronisation-process modification in presence of the colour-deconfined medium. 

The recent upgrade of the ALICE apparatus will greatly enhance the physics potential of the experiment in the LHC Run 3 data-taking period, allowing for more precise measurements of the non-prompt $\Ds$-meson production in heavy--ion collisions. 

%
%%%%%%%%%%%%%%%%%%%%%%%%%%%%%%%%
% end main text 
%%%%%%%%%%%%%%%%%%%%%%%%%%%%%%%%

%%%%% acknowledgements - handled by EB chairs 
\newenvironment{acknowledgement}{\relax}{\relax}
\begin{acknowledgement}
\section*{Acknowledgements}
%% add specific acknowledgements here 
%% ...but please don't remove the line below: funding agencies
%% will be acknowledged with a custom tex file handled by EB chairs after Collab Round 2
% Version: 2022-03-12

The ALICE Collaboration would like to thank all its engineers and technicians for their invaluable contributions to the construction of the experiment and the CERN accelerator teams for the outstanding performance of the LHC complex.
The ALICE Collaboration gratefully acknowledges the resources and support provided by all Grid centres and the Worldwide LHC Computing Grid (WLCG) collaboration.
The ALICE Collaboration acknowledges the following funding agencies for their support in building and running the ALICE detector:
A. I. Alikhanyan National Science Laboratory (Yerevan Physics Institute) Foundation (ANSL), State Committee of Science and World Federation of Scientists (WFS), Armenia;
Austrian Academy of Sciences, Austrian Science Fund (FWF): [M 2467-N36] and Nationalstiftung f\"{u}r Forschung, Technologie und Entwicklung, Austria;
Ministry of Communications and High Technologies, National Nuclear Research Center, Azerbaijan;
Conselho Nacional de Desenvolvimento Cient\'{\i}fico e Tecnol\'{o}gico (CNPq), Financiadora de Estudos e Projetos (Finep), Funda\c{c}\~{a}o de Amparo \`{a} Pesquisa do Estado de S\~{a}o Paulo (FAPESP) and Universidade Federal do Rio Grande do Sul (UFRGS), Brazil;
Bulgarian Ministry of Education and Science, within the National Roadmap for Research Infrastructures 2020-2027 (object CERN), Bulgaria;
Ministry of Education of China (MOEC) , Ministry of Science \& Technology of China (MSTC) and National Natural Science Foundation of China (NSFC), China;
Ministry of Science and Education and Croatian Science Foundation, Croatia;
Centro de Aplicaciones Tecnol\'{o}gicas y Desarrollo Nuclear (CEADEN), Cubaenerg\'{\i}a, Cuba;
Ministry of Education, Youth and Sports of the Czech Republic, Czech Republic;
The Danish Council for Independent Research | Natural Sciences, the VILLUM FONDEN and Danish National Research Foundation (DNRF), Denmark;
Helsinki Institute of Physics (HIP), Finland;
Commissariat \`{a} l'Energie Atomique (CEA) and Institut National de Physique Nucl\'{e}aire et de Physique des Particules (IN2P3) and Centre National de la Recherche Scientifique (CNRS), France;
Bundesministerium f\"{u}r Bildung und Forschung (BMBF) and GSI Helmholtzzentrum f\"{u}r Schwerionenforschung GmbH, Germany;
General Secretariat for Research and Technology, Ministry of Education, Research and Religions, Greece;
National Research, Development and Innovation Office, Hungary;
Department of Atomic Energy Government of India (DAE), Department of Science and Technology, Government of India (DST), University Grants Commission, Government of India (UGC) and Council of Scientific and Industrial Research (CSIR), India;
National Research and Innovation Agency - BRIN, Indonesia;
Istituto Nazionale di Fisica Nucleare (INFN), Italy;
Japanese Ministry of Education, Culture, Sports, Science and Technology (MEXT) and Japan Society for the Promotion of Science (JSPS) KAKENHI, Japan;
Consejo Nacional de Ciencia (CONACYT) y Tecnolog\'{i}a, through Fondo de Cooperaci\'{o}n Internacional en Ciencia y Tecnolog\'{i}a (FONCICYT) and Direcci\'{o}n General de Asuntos del Personal Academico (DGAPA), Mexico;
Nederlandse Organisatie voor Wetenschappelijk Onderzoek (NWO), Netherlands;
The Research Council of Norway, Norway;
Commission on Science and Technology for Sustainable Development in the South (COMSATS), Pakistan;
Pontificia Universidad Cat\'{o}lica del Per\'{u}, Peru;
Ministry of Education and Science, National Science Centre and WUT ID-UB, Poland;
Korea Institute of Science and Technology Information and National Research Foundation of Korea (NRF), Republic of Korea;
Ministry of Education and Scientific Research, Institute of Atomic Physics, Ministry of Research and Innovation and Institute of Atomic Physics and University Politehnica of Bucharest, Romania;
Ministry of Education, Science, Research and Sport of the Slovak Republic, Slovakia;
National Research Foundation of South Africa, South Africa;
Swedish Research Council (VR) and Knut \& Alice Wallenberg Foundation (KAW), Sweden;
European Organization for Nuclear Research, Switzerland;
Suranaree University of Technology (SUT), National Science and Technology Development Agency (NSTDA) and National Science, Research and Innovation Fund (NSRF via PMU-B B05F650021), Thailand;
Turkish Energy, Nuclear and Mineral Research Agency (TENMAK), Turkey;
National Academy of  Sciences of Ukraine, Ukraine;
Science and Technology Facilities Council (STFC), United Kingdom;
National Science Foundation of the United States of America (NSF) and United States Department of Energy, Office of Nuclear Physics (DOE NP), United States of America.
In addition, individual groups or members have received support from:
Marie Sk\l{}odowska Curie, Strong 2020 - Horizon 2020, European Research Council (grant nos. 824093, 896850, 950692), European Union;
Academy of Finland (Center of Excellence in Quark Matter) (grant nos. 346327, 346328), Finland;
Programa de Apoyos para la Superaci\'{o}n del Personal Acad\'{e}mico, UNAM, Mexico.
\end{acknowledgement}

%%%%%%%% Bibliography 
\bibliographystyle{utphys}   % Remember we use title in the biblio
\bibliography{bibliography}

%%%%%%%%%%%%%%%%%%%%%%%%%%%%%%%%
% Appendices: yours (if any) + authorlist
%%%%%%%%%%%%%%%%%%%%%%%%%%%%%%%%
\newpage
\appendix

%
%\input{} % put your appendices here (if any)
%

%%%%% Authorlist - please do not touch: handled by EB chairs 
\section{The ALICE Collaboration}
\label{app:collab}
% ALICE Collaboration author list for 2022-03-12
\begin{flushleft} 
\small

S.~Acharya\,\orcidlink{0000-0002-9213-5329}\,$^{\rm 123,131}$, 
D.~Adamov\'{a}\,\orcidlink{0000-0002-0504-7428}\,$^{\rm 85}$, 
A.~Adler$^{\rm 69}$, 
G.~Aglieri Rinella\,\orcidlink{0000-0002-9611-3696}\,$^{\rm 32}$, 
M.~Agnello\,\orcidlink{0000-0002-0760-5075}\,$^{\rm 29}$, 
N.~Agrawal\,\orcidlink{0000-0003-0348-9836}\,$^{\rm 50}$, 
Z.~Ahammed\,\orcidlink{0000-0001-5241-7412}\,$^{\rm 131}$, 
S.~Ahmad\,\orcidlink{0000-0003-0497-5705}\,$^{\rm 15}$, 
S.U.~Ahn\,\orcidlink{0000-0001-8847-489X}\,$^{\rm 70}$, 
I.~Ahuja\,\orcidlink{0000-0002-4417-1392}\,$^{\rm 37}$
% ...

\section*{Affiliation Notes}

$^{\rm I}$ Deceased\\
$^{\rm II}$ Also at: Max-Planck-Institut f\"{u}r Physik, Munich, Germany\\
$^{\rm III}$ Also at: Italian National Agency for New Technologies, Energy and Sustainable Economic Development (ENEA), Bologna, Italy\\
$^{\rm IV}$ Also at: Dipartimento DET del Politecnico di Torino, Turin, Italy\\
$^{\rm V}$ Also at: Department of Applied Physics, Aligarh Muslim University, Aligarh, India\\
$^{\rm VI}$ Also at: Institute of Theoretical Physics, University of Wroclaw, Poland\\
$^{\rm VII}$ Also at: An institution covered by a cooperation agreement with CERN\\

\section*{Collaboration Institutes}

$^{1}$ A.I. Alikhanyan National Science Laboratory (Yerevan Physics Institute) Foundation, Yerevan, Armenia\\
$^{2}$ AGH University of Krakow, Cracow, Poland\\
$^{3}$ Bogolyubov Institute for Theoretical Physics, National Academy of Sciences of Ukraine, Kiev, Ukraine\\
$^{4}$ Bose Institute, Department of Physics  and Centre for Astroparticle Physics and Space Science (CAPSS), Kolkata, India\\
$^{5}$ California Polytechnic State University, San Luis Obispo, California, United States\\
$^{6}$ Central China Normal University, Wuhan, China\\
$^{7}$ Centro de Aplicaciones Tecnol\'{o}gicas y Desarrollo Nuclear (CEADEN), Havana, Cuba\\
$^{8}$ Centro de Investigaci\'{o}n y de Estudios Avanzados (CINVESTAV), Mexico City and M\'{e}rida, Mexico\\
$^{9}$ Chicago State University, Chicago, Illinois, United States\\
$^{10}$ China Institute of Atomic Energy, Beijing, China\\
$^{11}$ Chungbuk National University, Cheongju, Republic of Korea\\
$^{12}$ Comenius University Bratislava, Faculty of Mathematics, Physics and Informatics, Bratislava, Slovak Republic\\
$^{13}$ COMSATS University Islamabad, Islamabad, Pakistan\\
$^{14}$ Creighton University, Omaha, Nebraska, United States\\
$^{15}$ Department of Physics, Aligarh Muslim University, Aligarh, India\\
$^{16}$ Department of Physics, Pusan National University, Pusan, Republic of Korea\\
$^{17}$ Department of Physics, Sejong University, Seoul, Republic of Korea\\
$^{18}$ Department of Physics, University of California, Berkeley, California, United States\\
$^{19}$ Department of Physics, University of Oslo, Oslo, Norway\\
$^{20}$ Department of Physics and Technology, University of Bergen, Bergen, Norway\\
$^{21}$ Dipartimento di Fisica, Universit\`{a} di Pavia, Pavia, Italy\\
$^{22}$ Dipartimento di Fisica dell'Universit\`{a} and Sezione INFN, Cagliari, Italy\\
$^{23}$ Dipartimento di Fisica dell'Universit\`{a} and Sezione INFN, Trieste, Italy\\
$^{24}$ Dipartimento di Fisica dell'Universit\`{a} and Sezione INFN, Turin, Italy\\
$^{25}$ Dipartimento di Fisica e Astronomia dell'Universit\`{a} and Sezione INFN, Bologna, Italy\\
$^{26}$ Dipartimento di Fisica e Astronomia dell'Universit\`{a} and Sezione INFN, Catania, Italy\\
$^{27}$ Dipartimento di Fisica e Astronomia dell'Universit\`{a} and Sezione INFN, Padova, Italy\\
$^{28}$ Dipartimento di Fisica `E.R.~Caianiello' dell'Universit\`{a} and Gruppo Collegato INFN, Salerno, Italy\\
$^{29}$ Dipartimento DISAT del Politecnico and Sezione INFN, Turin, Italy\\
$^{30}$ Dipartimento di Scienze MIFT, Universit\`{a} di Messina, Messina, Italy\\
$^{31}$ Dipartimento Interateneo di Fisica `M.~Merlin' and Sezione INFN, Bari, Italy\\
$^{32}$ European Organization for Nuclear Research (CERN), Geneva, Switzerland\\
$^{33}$ Faculty of Electrical Engineering, Mechanical Engineering and Naval Architecture, University of Split, Split, Croatia\\
$^{34}$ Faculty of Engineering and Science, Western Norway University of Applied Sciences, Bergen, Norway\\
$^{35}$ Faculty of Nuclear Sciences and Physical Engineering, Czech Technical University in Prague, Prague, Czech Republic\\
$^{36}$ Faculty of Physics, Sofia University, Sofia, Bulgaria\\
$^{37}$ Faculty of Science, P.J.~\v{S}af\'{a}rik University, Ko\v{s}ice, Slovak Republic\\
$^{38}$ Frankfurt Institute for Advanced Studies, Johann Wolfgang Goethe-Universit\"{a}t Frankfurt, Frankfurt, Germany\\
$^{39}$ Fudan University, Shanghai, China\\
$^{40}$ Gangneung-Wonju National University, Gangneung, Republic of Korea\\
$^{41}$ Gauhati University, Department of Physics, Guwahati, India\\
$^{42}$ Helmholtz-Institut f\"{u}r Strahlen- und Kernphysik, Rheinische Friedrich-Wilhelms-Universit\"{a}t Bonn, Bonn, Germany\\
$^{43}$ Helsinki Institute of Physics (HIP), Helsinki, Finland\\
$^{44}$ High Energy Physics Group,  Universidad Aut\'{o}noma de Puebla, Puebla, Mexico\\
$^{45}$ Horia Hulubei National Institute of Physics and Nuclear Engineering, Bucharest, Romania\\
$^{46}$ Indian Institute of Technology Bombay (IIT), Mumbai, India\\
$^{47}$ Indian Institute of Technology Indore, Indore, India\\
$^{48}$ INFN, Laboratori Nazionali di Frascati, Frascati, Italy\\
$^{49}$ INFN, Sezione di Bari, Bari, Italy\\
$^{50}$ INFN, Sezione di Bologna, Bologna, Italy\\
$^{51}$ INFN, Sezione di Cagliari, Cagliari, Italy\\
$^{52}$ INFN, Sezione di Catania, Catania, Italy\\
$^{53}$ INFN, Sezione di Padova, Padova, Italy\\
$^{54}$ INFN, Sezione di Pavia, Pavia, Italy\\
$^{55}$ INFN, Sezione di Torino, Turin, Italy\\
$^{56}$ INFN, Sezione di Trieste, Trieste, Italy\\
$^{57}$ Inha University, Incheon, Republic of Korea\\
$^{58}$ Institute for Gravitational and Subatomic Physics (GRASP), Utrecht University/Nikhef, Utrecht, Netherlands\\
$^{59}$ Institute of Experimental Physics, Slovak Academy of Sciences, Ko\v{s}ice, Slovak Republic\\
$^{60}$ Institute of Physics, Homi Bhabha National Institute, Bhubaneswar, India\\
$^{61}$ Institute of Physics of the Czech Academy of Sciences, Prague, Czech Republic\\
$^{62}$ Institute of Space Science (ISS), Bucharest, Romania\\
$^{63}$ Institut f\"{u}r Kernphysik, Johann Wolfgang Goethe-Universit\"{a}t Frankfurt, Frankfurt, Germany\\
$^{64}$ Instituto de Ciencias Nucleares, Universidad Nacional Aut\'{o}noma de M\'{e}xico, Mexico City, Mexico\\
$^{65}$ Instituto de F\'{i}sica, Universidade Federal do Rio Grande do Sul (UFRGS), Porto Alegre, Brazil\\
$^{66}$ Instituto de F\'{\i}sica, Universidad Nacional Aut\'{o}noma de M\'{e}xico, Mexico City, Mexico\\
$^{67}$ iThemba LABS, National Research Foundation, Somerset West, South Africa\\
$^{68}$ Jeonbuk National University, Jeonju, Republic of Korea\\
$^{69}$ Johann-Wolfgang-Goethe Universit\"{a}t Frankfurt Institut f\"{u}r Informatik, Fachbereich Informatik und Mathematik, Frankfurt, Germany\\
$^{70}$ Korea Institute of Science and Technology Information, Daejeon, Republic of Korea\\
$^{71}$ KTO Karatay University, Konya, Turkey\\
$^{72}$ Laboratoire de Physique Subatomique et de Cosmologie, Universit\'{e} Grenoble-Alpes, CNRS-IN2P3, Grenoble, France\\
$^{73}$ Lawrence Berkeley National Laboratory, Berkeley, California, United States\\
$^{74}$ Lund University Department of Physics, Division of Particle Physics, Lund, Sweden\\
$^{75}$ Nagasaki Institute of Applied Science, Nagasaki, Japan\\
$^{76}$ Nara Women{'}s University (NWU), Nara, Japan\\
$^{77}$ National and Kapodistrian University of Athens, School of Science, Department of Physics , Athens, Greece\\
$^{78}$ National Centre for Nuclear Research, Warsaw, Poland\\
$^{79}$ National Institute of Science Education and Research, Homi Bhabha National Institute, Jatni, India\\
$^{80}$ National Nuclear Research Center, Baku, Azerbaijan\\
$^{81}$ National Research and Innovation Agency - BRIN, Jakarta, Indonesia\\
$^{82}$ Niels Bohr Institute, University of Copenhagen, Copenhagen, Denmark\\
$^{83}$ Nikhef, National institute for subatomic physics, Amsterdam, Netherlands\\
$^{84}$ Nuclear Physics Group, STFC Daresbury Laboratory, Daresbury, United Kingdom\\
$^{85}$ Nuclear Physics Institute of the Czech Academy of Sciences, Husinec-\v{R}e\v{z}, Czech Republic\\
$^{86}$ Oak Ridge National Laboratory, Oak Ridge, Tennessee, United States\\
$^{87}$ Ohio State University, Columbus, Ohio, United States\\
$^{88}$ Physics department, Faculty of science, University of Zagreb, Zagreb, Croatia\\
$^{89}$ Physics Department, Panjab University, Chandigarh, India\\
$^{90}$ Physics Department, University of Jammu, Jammu, India\\
$^{91}$ Physics Department, University of Rajasthan, Jaipur, India\\
$^{92}$ Physics Program and International Institute for Sustainability with Knotted Chiral Meta Matter (SKCM2), Hiroshima University, Hiroshima, Japan\\
$^{93}$ Physikalisches Institut, Eberhard-Karls-Universit\"{a}t T\"{u}bingen, T\"{u}bingen, Germany\\
$^{94}$ Physikalisches Institut, Ruprecht-Karls-Universit\"{a}t Heidelberg, Heidelberg, Germany\\
$^{95}$ Physik Department, Technische Universit\"{a}t M\"{u}nchen, Munich, Germany\\
$^{96}$ Politecnico di Bari and Sezione INFN, Bari, Italy\\
$^{97}$ Research Division and ExtreMe Matter Institute EMMI, GSI Helmholtzzentrum f\"ur Schwerionenforschung GmbH, Darmstadt, Germany\\
$^{98}$ Saha Institute of Nuclear Physics, Homi Bhabha National Institute, Kolkata, India\\
$^{99}$ School of Physics and Astronomy, University of Birmingham, Birmingham, United Kingdom\\
$^{100}$ Secci\'{o}n F\'{\i}sica, Departamento de Ciencias, Pontificia Universidad Cat\'{o}lica del Per\'{u}, Lima, Peru\\
$^{101}$ Stefan Meyer Institut f\"{u}r Subatomare Physik (SMI), Vienna, Austria\\
$^{102}$ SUBATECH, IMT Atlantique, Nantes Universit\'{e}, CNRS-IN2P3, Nantes, France\\
$^{103}$ Suranaree University of Technology, Nakhon Ratchasima, Thailand\\
$^{104}$ Technical University of Ko\v{s}ice, Ko\v{s}ice, Slovak Republic\\
$^{105}$ The Henryk Niewodniczanski Institute of Nuclear Physics, Polish Academy of Sciences, Cracow, Poland\\
$^{106}$ The University of Texas at Austin, Austin, Texas, United States\\
$^{107}$ Universidad Aut\'{o}noma de Sinaloa, Culiac\'{a}n, Mexico\\
$^{108}$ Universidade de S\~{a}o Paulo (USP), S\~{a}o Paulo, Brazil\\
$^{109}$ Universidade Estadual de Campinas (UNICAMP), Campinas, Brazil\\
$^{110}$ Universidade Federal do ABC, Santo Andre, Brazil\\
$^{111}$ University of Cape Town, Cape Town, South Africa\\
$^{112}$ University of Houston, Houston, Texas, United States\\
$^{113}$ University of Jyv\"{a}skyl\"{a}, Jyv\"{a}skyl\"{a}, Finland\\
$^{114}$ University of Kansas, Lawrence, Kansas, United States\\
$^{115}$ University of Liverpool, Liverpool, United Kingdom\\
$^{116}$ University of Science and Technology of China, Hefei, China\\
$^{117}$ University of South-Eastern Norway, Kongsberg, Norway\\
$^{118}$ University of Tennessee, Knoxville, Tennessee, United States\\
$^{119}$ University of the Witwatersrand, Johannesburg, South Africa\\
$^{120}$ University of Tokyo, Tokyo, Japan\\
$^{121}$ University of Tsukuba, Tsukuba, Japan\\
$^{122}$ University Politehnica of Bucharest, Bucharest, Romania\\
$^{123}$ Universit\'{e} Clermont Auvergne, CNRS/IN2P3, LPC, Clermont-Ferrand, France\\
$^{124}$ Universit\'{e} de Lyon, CNRS/IN2P3, Institut de Physique des 2 Infinis de Lyon, Lyon, France\\
$^{125}$ Universit\'{e} de Strasbourg, CNRS, IPHC UMR 7178, F-67000 Strasbourg, France, Strasbourg, France\\
$^{126}$ Universit\'{e} Paris-Saclay, Centre d'Etudes de Saclay (CEA), IRFU, D\'{e}partment de Physique Nucl\'{e}aire (DPhN), Saclay, France\\
$^{127}$ Universit\'{e}  Paris-Saclay, CNRS/IN2P3, IJCLab, Orsay, France\\
$^{128}$ Universit\`{a} degli Studi di Foggia, Foggia, Italy\\
$^{129}$ Universit\`{a} del Piemonte Orientale, Vercelli, Italy\\
$^{130}$ Universit\`{a} di Brescia, Brescia, Italy\\
$^{131}$ Variable Energy Cyclotron Centre, Homi Bhabha National Institute, Kolkata, India\\
$^{132}$ Warsaw University of Technology, Warsaw, Poland\\
$^{133}$ Wayne State University, Detroit, Michigan, United States\\
$^{134}$ Westf\"{a}lische Wilhelms-Universit\"{a}t M\"{u}nster, Institut f\"{u}r Kernphysik, M\"{u}nster, Germany\\
$^{135}$ Wigner Research Centre for Physics, Budapest, Hungary\\
$^{136}$ Yale University, New Haven, Connecticut, United States\\
$^{137}$ Yonsei University, Seoul, Republic of Korea\\
$^{138}$  Zentrum  f\"{u}r Technologie und Transfer (ZTT), Worms, Germany\\
$^{139}$ Affiliated with an institute covered by a cooperation agreement with CERN\\
$^{140}$ Affiliated with an international laboratory covered by a cooperation agreement with CERN.\\

\end{flushleft} 

\end{document}